\documentclass{emulateapj}

\usepackage{natbib}
\usepackage{longtable,lscape}
\begin{document}

\title{The stellar activity--rotation relationship \\ and the evolution of stellar dynamos}
\author{Nicholas J. Wright$^1$, Jeremy J. Drake$^1$, Eric E. Mamajek$^2$, and Gregory W. Henry$^3$}
\affil{$^1$ Harvard-Smithsonian Center for Astrophysics, 60 Garden Street, Cambridge, MA~02138.\\
$^2$ Department of Physics and Astronomy, University of Rochester, Rochester, NY~14627.\\
$^3$ Center of Excellence in Information Systems, Tennessee State University, 3500 John A. Merritt Blvd., Box 9501, Nashville, TN~37209.}
\email{nwright@cfa.harvard.edu}

\begin{abstract}

We present a sample of 824 solar and late-type stars with X-ray luminosities and rotation periods. This is used to study the relationship between rotation and stellar activity and derive a new estimate of the convective turnover time. From an unbiased subset of this sample the power law slope of the unsaturated regime, $L_X/L_{bol}\propto Ro^\beta$, is fit as $\beta=-2.70\pm0.13$. This is inconsistent with the canonical $\beta=-2$ slope to a confidence of 5$\sigma$, and argues for an additional term in the dynamo number equation. From a simple scaling analysis this implies $\Delta\Omega/\Omega\propto\Omega^{0.7}$, i.e. the differential rotation of solar-type stars gradually declines as they spin down. Super-saturation is observed for the fastest rotators in our sample and its parametric dependencies are explored. Significant correlations are found with both the corotation radius and the excess polar updraft, the latter theory providing a stronger dependence and being supported by other observations. We estimate mass-dependent empirical thresholds for saturation and super-saturation and map out three regimes of coronal emission. Late F-type stars are shown never to pass through the saturated regime, passing straight from super-saturated to unsaturated X-ray emission. The theoretical threshold for coronal stripping is shown to be significantly different from the empirical saturation threshold ($Ro < 0.13$), suggesting it is not responsible. Instead we suggest that a different dynamo configuration is at work in stars with saturated coronal emission. This is supported by a correlation between the empirical saturation threshold and the time when stars transition between convective and interface sequences in rotational spin-down models.

\end{abstract}


\section{Introduction}

Stars across the Hertzsprung-Russell diagram are known to emit X-rays with only a few exceptions. The hottest and most massive stars emit X-rays that arise from either small-scale shocks in their winds \citep{lucy80} or collisions between the wind and circumstellar material \citep[e.g.][]{park09}. Solar and late-type stars emit X-rays from a magnetically confined plasma at temperatures of several million Kelvin known as a corona \citep{vaia81}. This high-temperature rarefied thermal plasma, first observed on the Sun, has been detected from nearly all classes of low-mass stars at levels of $L_X / L_{bol} \sim 10^{-8} - 10^{-3}$ \citep{schm04}. The most X-ray luminous of single, low-mass stars have been found to be the youngest, where X-ray luminosity levels of $L_X / L_{bol} \sim 10^{-3}$ are common \citep[e.g.][]{tell07}, though close or interacting binary stars such as RS~CVns may also reach such levels. Older stars have typically lower X-ray luminosities, reaching down to $L_X / L_{bol} \sim 10^{-8} - 10^{-4}$ \citep[e.g.][]{feig04,wrig10b}. This significant decrease in stellar X-ray luminosity of several orders of magnitude between the zero age main sequence and the solar age is widely used to identify young stars efficiently against the Galactic field population \citep[e.g.][]{getm05,wrig09a}.

Coronal X-ray emission, as well as other indicators of stellar activity \citep[e.g. Ca~{\sc ii} or H$\alpha$ emission,][]{west08}, is believed to be driven by the stellar magnetic dynamo, which itself is thought to be driven by differential rotation in the stellar interior \citep[e.g.][]{park55,wils66,kraf67}, a phenomena that has been confirmed in the Sun through helioseismology \citep[e.g.][]{duva84}. The observed decrease in X-ray emission over the lifetime of the star can therefore be attributed to the rotational spin-down of the star, which is driven by mass-loss through a magnetized stellar wind \citep[e.g.][]{skum72}. Because the magnetized stellar wind is also believed to be driven by the stellar dynamo \citep[e.g.][]{cran11}, rotation and magnetic activity effectively operate in a feedback loop with rotation, stellar activity, and the stellar wind decreasing with age. Since angular momentum loss controls the evolution of close binaries that are believed to be the progenitors of cataclysmic variables, novae, and Type~1a supernovae, understanding these dependencies has far-reaching implications.

The relationship between stellar rotation and tracers of magnetic activity is an important probe of the physical dynamo process, with chromospheric and coronal tracers increasing monotonically with increasing rotational velocity for periods exceeding a few days. A relationship between rotation and activity was first quantified by \citet{pall81} who found that X-ray luminosity scaled as $L_X \propto (v \mathrm{sin} i)^{1.9}$, providing the first evidence for the dynamo-induced nature of stellar coronal activity. For very fast rotators the relationship was found to break down \citep{mice85}, with X-ray luminosity reaching a saturation level of approximately log~$L_X / L_{bol} \sim -3$ \citep{vilh84,vilh87}, independent of spectral-type. This saturation level is reached at a rotation period that increases toward later spectral types \citep[increasing with decreasing bolometric luminosity,][]{pizz03}. It is unclear whether this saturation is caused by a saturation of the dynamo itself \citep[e.g.][]{vilh84}, a saturation of the filling factor of active regions on the stellar surface \citep{vilh84}, or a centrifugal stripping of the corona caused by the high rotation rates \citep{jard99}. However, once saturation occurs the X-ray emission becomes a function of only the bolometric luminosity \citep{pizz03}, or effectively the mass, color or radius of the main-sequence star.

In the non-saturated regime, the two influences on the efficiency of the magnetic dynamo were combined by \citet{noye84} into a single parameter, the Rossby number, $R_0 = P_{rot} / \tau$, the ratio of the stellar rotation period, $P_{rot}$, and the  mass-dependent convective turnover time, $\tau$. This quantity has proven to be an effective parameter of the stellar magnetic dynamo, increasing toward lower masses with the efficiency of the dynamo \citep[e.g.][]{mice84,magg87,step94,rand00}. Despite this work there is yet to be a satisfactory dynamo theory that can explain both the solar dynamo and that of rapidly rotating stars \citep[e.g.][]{weis05,bran11} and the continued lack of a sufficiently large and unbiased sample has no doubt contributed to this.

The paucity of stellar samples with which to study the rotation--activity relationship has mainly been due to the difficulty of measuring accurate stellar rotation periods, which require multiple deep observations over long baselines. This has led to the use of projected rotational velocities as a substitute, which are influenced by the uncertainties of estimated stellar radii and unknown inclination angles. The recent increase in measured rotation periods \citep[e.g.][increased the number of Pleiades stars with measured periods by a factor of five]{hart10} for many thousands of stars in open clusters of known age is overcoming this problem and it is likely that we will henceforth be limited by the availability of deep X-ray observations for such stars.

In this work, we combine new measurements of photometric rotation periods for a large number of field and cluster stars with archival X-ray observations to produce the largest existing sample of stars with photometric rotation periods and X-ray luminosities (Section~2). This sample is then used in Section~3 to study and characterize the rotation - activity relationship in detail and to probe the stellar magnetic dynamo responsible for it. This allows us in Section~4 to trace out the X-ray evolution of low-mass stars as a function of rotation period, which is a good proxy for age. Finally, in Section~5 this sample is used to derive a new empirical measure of the mass-dependent convective turnover time.

\section{Compilation of the sample}

To study the relationship between rotation and activity a sample was compiled from the literature by searching for stars with measurements of both rotation periods and X-ray luminosities. Only photometrically-determined rotation periods were included, discarding all rotation velocity measurements and upper limits, and only stars with significant X-ray detections were used, discarding all sources with only upper limits. This choice reduces the sample size available and also has the potential to introduce an X-ray luminosity bias in our results. However this greatly simplifies the following analysis, particularly in the light of the large variety of sources used to compile this sample, the different techniques used to calculate upper limits by different authors, and the potential incompletenesses in upper limits present in each sample. The inherent biases that will exist in this sample will be discussed and addressed later.

\subsection{Rotation periods and X-ray fluxes}

The recent study of the activity--rotation relation by \citet{pizz03} provided the starting point for the catalog. From their work 102 cluster stars (excluding Pleiades members, which were compiled separately) and 47 field stars were used, excluding all sources with upper limits, as well as a number of stars whose rotation periods were inferred indirectly from chromospheric activity levels (with the exception of $\alpha$~Centauri~B, for which we use the rotation period and mean X-ray luminosity presented by \citet{dewa10}). The majority of these sources are G and K stars with the M-type stars confined to the field star sample because of the lack of rotational periods available for low mass stars in clusters at the time. A further 28 Hyades members were introduced by cross-matching recent rotation periods from \citet{delo11} with X-ray luminosities from \citet{ster94} and \citet{ster95}.

The first expansion of the sample was based on the recent measurement of photometric rotation periods by \citet{hart10} for Pleiades members based on the membership list of \citet{stau07}. These were cross-matched with X-ray flux measurements of Pleiades stars from the {\it Einstein} and {\it ROSAT} observations from \citet{mice90}, \citet{stau94}, \citet{mice96}, and \citet{mice99}, and {\it XMM-Newton} observations from \citet{brig03}, again discarding all upper limits. Where multiple measurements of the X-ray flux exist for a single source, that with the lowest fractional uncertainty was used. Cross-matching 391 X-ray sources with the 383 sources with photometric periods lead to a sample of 146 Pleiades members with both X-ray luminosities and rotation periods, the largest such sample for a single cluster.

This sample was then complemented with 83 stars from the open clusters NGC~2516 and NGC~2547 using, respectively, rotation periods from \citet{irwi07} and \citet{irwi08}, and X-ray fluxes from \citet{pill06} and \citet{jeff11}. An additional 20 stars were added from the open cluster Praesepe using rotation periods from \citet{delo11} and \citet{scho11} and X-ray luminosities from \citet{rand95} and \citet{fran03}. The sample was then further extended using data from rotation period surveys of field stars combined with X-ray fluxes from the {\it ROSAT} All-Sky Survey. This included 218 stars from \citet{hart11}, 23 stars from \citet{kira07}, 6 stars from \citet{xing07}, 8 stars from \citet{bouv97} and 79 stars were added from the compilation of \citet{mama08}. This sample also includes the Sun using the values of log~$L_X / L_{bol} = -6.24$ \citep{judg03}\footnote{This value is larger than that of \citet{pere00} by a factor two, a factor that would only exacerbates the poor fit of the solar data point compared to the fit in Section~\ref{s-unsat}.} and $P_{rot} = 26.09$~days \citep{dona96}. Finally, rotation periods for 65 stars were taken from observations as part of the FEPS (Formation and Evolution of Planetary Systems, see Appendix~A) program.

Known classical T-Tauri stars with H$\alpha$ emission, signaling the presence of accretion and therefore a circumstellar disk, were excluded because of the complications induced by X-ray emission from accretion and disk-locking on the rotation period. Pre-MS stars (specifically those with ages $<$10~Myrs) were also excluded because of potential differences in their internal structure as a function of either mass or effective temperature. Using the catalog of X-ray variable sources presented by \citet{fuhr03}, we removed all ROSAT sources that had been observed to flare during the observation to lessen the influence of short-duration flares on the resulting X-ray luminosities. In total the resulting sample includes 824 stars, all with photometrically-measured rotation periods and well-constrained X-ray flux measurements.

\subsection{Homogenizing the sample}

The majority of previous work on the rotation--activity relation has used the $B-V$ color as the proxy for effective temperature. This was due primarily to the ubiquity of suitable photometry for the FGK stars that made up these studies and the lack of photometry in other bands. However, the increased interest in M-type stars, for which the $B-V$ color is a poor temperature diagnostic, and the availability of near-IR photometry suggests an alternative color as an appropriate proxy. The $V-K_s$ color, derived from now readily available optical and near-IR photometry offers a sufficiently large baseline such that uncertainties in either individual magnitude do not greatly affect the derived color. Complementary photometry was therefore sought from the Hipparcos \citep{perr97} and Two Micron All Sky Survey \citep[2MASS,][]{cutr03} catalogs and used to derive $V-K_s$ colors for all stars from the available photometric colors with the widest baseline, using the empirical tabulations of $V-K_s$ versus other colors and spectral type presented by \citet{peca11}. 797 sources have $V$ and near-IR photometry (2MASS $J$, $H$, or $K_s$ band observations with the lowest photometric uncertainty and photometric quality flags of A, B, or C), 17 sources had $V-I$ colors from which $V-K_s$ colors were derived, and 10 sources had only $B-V$ photometry, due primarily to their proximity and the saturation of near-IR photometry. The literature was then searched for suitable spectral types for these sources (see Appendix~B for references) and these were used to derive the intrinsic $V-K_s$ color \citep[using the empirical tables of][]{peca11} for sources where the spectra would provide a more accurate diagnosis of the underlying photospheric temperature than photometry could\footnote{Given a typical classification uncertainty of $\pm$1 subtype we estimate that spectra provide a more accurate classification than photometry for all F and G-type stars, for stars known to be in binary systems, and for any source where the $B-V$ color was used. For late-type stars where either the $V-K_s$ or $V-I$ color was available we consider the photometry to have a smaller uncertainty than a typical spectral classification.}. In total 225 sources had spectral types in the literature that were used to derive dereddened $V-K_s$ colors, including all but 4 of the stars that lacked $V-K_s$ photometry. This also included all field stars more distant than 75~pc for which interstellar reddening might have been a concern for purely photometric data. The colors of cluster stars were dereddened using the best available measurements of cluster extinction (Table~\ref{sources}) to provide the most accurate available effective temperature proxy for each star in the sample. 

\begin{table}
\begin{center}
\caption{List of cluster and field star samples used in this work.} 
\label{sources}
\begin{tabular}{lcccc}
\tableline 
Cluster / field & Age & Distance & Extinction & Number \\
star sample & (Myrs) & (pc) & ($A_V$) & of stars \\
\tableline
NGC 2547$^a$	& 40			& 407	& 0.15	& 69 \\
IC 2602$^b$		& 46			& 149	& 0.09	& 28 \\
IC 2391$^c$		& 50			& 145	& 0.03	& 13 \\
$\alpha$ Persei$^c$& 85			& 172	& 0.27	& 40 \\
Pleiades$^d$		& 125		& 133	& 0.12	& 146 \\
NGC 2516$^e$	& 150		& 376	& 0.36	& 14 \\
Praesepe			& 600		& 182	& 0.08	& 20 \\
Hyades			& 700		& 46.5	& 0.0		& 49 \\
\tableline
Bouvier 1997		& -		& 143 - 264	&-		& 8 \\
Pizzolato 2003		& -		& 1.3 - 172	&-		& 47 \\
Kiraga 2007		& -		& 3.0 - 50		&-		& 23 \\
Xing 2007			& -		& 44 - 207		&-		& 6 \\
Mamajek 2008		& -		& 4 - 102		&-		& 79 \\
FEPS program		& -		& 19 - 102		&-		& 64 \\
Hartman 2011		& -		& 1.5 - 243	& -		& 218 \\
\tableline
Total & & & & 824 \\
\tableline
\end{tabular} 
\newline
Cluster distances and extinctions are taken from the Hipparcos parallax measurements \cite[][and references therein]{vanl09}, with the exception of the Pleiades, for which a distance of 133~pc is used \citep{stau98}, and the relatively distant clusters NGC~2516 and NGC~2547, which are poorly sampled by Hipparcos. Cluster ages are cited individually.\\
{\bf References.}
(a) \citet{mayn08}; 
(b) \citet{dobb10}; 
(c) \citet{barr04}; 
(d) \citet{stau98}; 
(e) \citet{jeff01}.
\end{center}
\end{table}

Using the $V-K_s$ color as a proxy for effective temperature, we derived bolometric corrections (BC) for all stars using the tabulation of \citet{keny95}\footnote{Conversions provided by \citet{bess05} were used to convert the $V-K$ color of this work onto the 2MASS-based $V-K_s$ color.}. Distances for field stars were calculated using Hipparcos \citep{perr97,vanl07} parallaxes, considering only those with good quality measurements, i.e. relative uncertainties $< 20$\%. For the 559 stars with known distances (378 cluster and 181 field stars) we combined these distances with BCs and $V$-band magnitudes to derive absolute bolometric luminosities. Combining the isochrones from \citet{sies00} and the most recent estimates for the ages of each cluster (Table~\ref{sources}), or assuming an age of 1~Gyr for field stars, we used the bolometric luminosities to derive stellar masses, radii, and effective temperatures for each star.

For the field stars for which distances were not known the $V-K_s$ color was used to derive the effective temperature using the calibrated effective temperature scale from \citet[][for M stars]{casa08} and \citet[][for FGK stars]{casa10}, with a linear interpolation between the two tabulations. The effective temperature was then fitted to the 1~Gyr \citet{sies00} main sequence isochrone to derive stellar masses, radii, bolometric luminosities, and therefore distances for all these stars. A comparison of the different color-T$_{eff}$ tabulations suggests that the systematic uncertainty in the derived stellar mass using this method is potentially as much as 10\% for FGK stars, and 25\% for M-type stars, though we are confident we have chosen the most recent and reliable color-T$_{eff}$ conversion for this use. For comparison, the derived mass uncertainty based on the color-BC conversion is $<2$\% for all stars with $V-K_s < 6$ (the majority of stars in our sample), excluding uncertainties in the distance used, the age of the cluster or systematic errors in the evolutionary tracks \citep[see e.g.][for a discussion of the accuracy of these tracks]{hill04}.

Using the derived distances, X-ray luminosities and X-ray-to-bolometric luminosity ratios were determined for all sources. For the majority of stars X-ray fluxes were available in the literature (based on X-ray spectral fits, count rates or hardness ratios), though for a small number of stars X-ray fluxes were either not available or had been calculated using a constant count-rate to flux conversion (i.e. one that does not take into account the temperature of the underlying spectrum through hardness ratios or spectral fits). This was mainly the case for some of the early ROSAT papers such as those studying the Hyades. For these sources we adopted the conversion between ROSAT count rates \citep{voge99} and hardness ratios published by \citet{flem95}. For comparison with the majority of literature X-ray observations of open clusters we have converted all X-ray luminosities to the ROSAT 0.1--2.4~keV energy band. This was done using PIMMS\footnote{The {\it Chandra} Portable Interactive Multi-Mission Simulator, http://cxc.harvard.edu/toolkit/pimms.jsp.}, assuming a thermal spectrum \citep{raym77} and a hydrogen column density converted from the visual extinction \citep{ryte96}. The temperature of the thermal spectrum for the conversion was chosen based on the activity level of the star, $L_X / L_{bol}$, and the observed correlation between activity level and plasma temperature \citep[see e.g.][]{tell05,gude09}\footnote{We calibrated this relation using the investigation of solar analogs by \citet{tell05}, but convert the dependency from the X-ray luminosity to the X-ray luminosity ratio as a more appropriate diagnostic of the activity level of the star.}. Uncertainties in these luminosities due to uncertainties in the hydrogen column density and plasma temperature are $\sim$5\% and $\sim$9\% for dispersions of $\Delta$~log~$N_H = 0.5$ and $\Delta kT = 0.5$~keV, as appropriate for the Pleiades \citep{gagn95}. For reference with other works the conversion factors to go from the ROSAT band to the {\it Chandra} 0.5--8.0~keV or {\it XMM-Newton} 0.3--4.5~keV bands are 0.676 and 0.797, respectively, for an active star with $kT \sim 1$~keV.

\subsection{The final catalog}

\begin{figure}
\begin{center}
\includegraphics[height=240pt, angle=270]{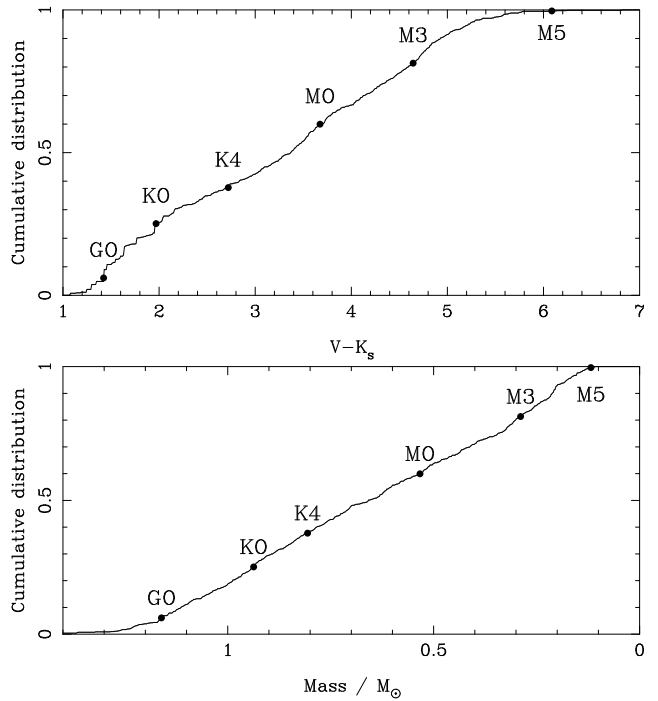}
\caption{Cumulative distribution of stars in our sample as a function of their $V-K$ color (top), used in this work as the observable proxy for effective temperature, and stellar mass (bottom), derived from the models of \citet{sies00}. Reference spectral types are indicated.}
\label{sample}
\end{center}
\end{figure}

The compilation of this data led to a catalog of 824 stars with rotation periods and X-ray luminosities that represents a significant increase over those used for previous studies of the activity--rotation relation by \citet[][$\sim$70 stars]{step94} and \citet[][$\sim$250 stars]{pizz03}. Table~\ref{sources} provides the details of all of these samples and the values used to derive the stellar parameters. Figure~\ref{sample} shows the distribution of the sample as a function of $V-K_s$ and mass. In $V-K_s$ color space the sample is approximately equally distributed across the range 1.5 - 5.0 (G2 - M4), with $\sim$30 stars per spectral subclass, dropping to $\sim$10 stars per subclass beyond these limits. Uncertainties of 10\% in the source or cluster distances translate to uncertainties $<$10\% in the resulting masses and uncertainties of $\sim$20\% in the resulting X-ray luminosities. However, distance uncertainties in the X-ray to bolometric luminosity ratios cancel out, and only uncertainties in the stellar mass become relevant. These uncertainties are generally small compared to uncertainties in the measured X-ray luminosities, which are found to have a typical uncertainty of $\sim$0.2~dex.

\section{X-ray emission vs. stellar rotation}
\label{s-main}

\begin{figure*}
\begin{center}
\includegraphics[height=510pt, angle=270]{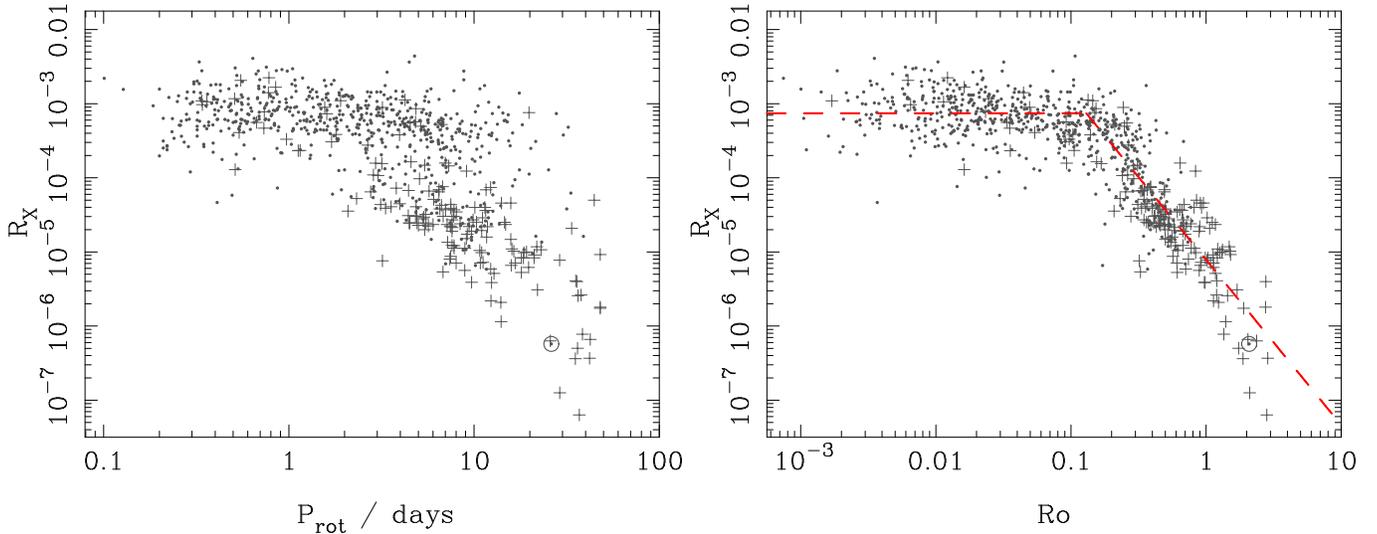}
\caption{X-ray to bolometric luminosity ratio plotted against rotation period (left panel) and the Rossby number, $Ro = P_{rot} / \tau$ (right panel), for all stars in our sample with X-ray luminosities and photometric rotation periods. Stars known to be binaries are shown as plus symbols, and the Sun is indicated with a solar symbol. The best-fitting saturated and non-saturated activity--rotation relations described in the text are shown as a dashed red line in the right-hand panel.}
\label{allstars}
\end{center}
\end{figure*}

In this section the accumulated data is used to study the relation between the level of X-ray activity, $L_X$, and the stellar rotation period, $P$, both of which vary by many orders of magnitude across the sample. These two parameters are thought to be connected by the magnetic dynamo \citep[e.g.][]{park55}, but the form of this relation and influence of spectral type is poorly understood. Because of this it has become usual to scale both observable parameters by functions of the stellar mass that allow a more useful comparison. \citet{pall81} represented the level of activity with the X-ray to bolometric luminosity ratio, $R_X = L_X / L_{bol}$. Following \citet{noye84} the rotation rate is represented with the Rossby number, $Ro = P / \tau$, the ratio of the stellar rotation period to the convective turnover time, $\tau$. The objective of these transformations is to convert the observable quantities into those that represent the parameters and products of the stellar dynamo. Figure~\ref{allstars} shows the X-ray luminosity ratio as a function of both the rotation period and the Rossby number and it is clear that the latter parameterization greatly reduces the scatter in the unsaturated regime.

The color-dependent convective turnover time, $\tau$, cannot be measured directly but can either be derived from models of stellar interiors \citep[e.g.][]{kim96} or be empirically estimated \citep[e.g.][]{noye84}. The semi-empirical determination provided by \citet{noye84} is the most well-used in the literature, but was derived from a small sample that only extends to $B-V = 1.4$, and as noted by \citet{pizz03} is based on only 5 points redder than $B-V = 1.0$. The empirically-derived values from \citet{pizz03} offer an improvement on this, being based on a sample that extends to $B-V = 2.0$, with a greater coverage of low-mass stars, and it is therefore chosen for this work. We use an empirically-derived conversion between $V-K_s$ and $B-V$ colors \citep{peca11} to estimate convective turnover times for each star. We note that the convective turnover time is known to vary over the life of a star \citep[e.g.][]{kim96} and may vary significantly in the pre-MS phase, potentially causing a systematic error in the analysis of the rotation--activity relation for such stars. In Section~\ref{s-tau} we follow the method of \citet{pizz03} to derive a new empirical estimate of $\tau$ based on the larger sample used in this work.

Figure~\ref{allstars} shows $R_X$ as a function of $Ro$ for all the stars in our sample, clearly demonstrating that the X-ray luminosity ratio increases with decreasing rotation period (or increasing rotation rate), as expected from qualitative arguments based on the $\alpha-\omega$-type shell dynamo theory. As noted by many previous observers, the X-ray emission level appears to saturate at the highest rotation rates, reaching an approximately constant level of $R_X \sim 10^{-3}$. This effect clearly divides the rotation--activity relation into two regimes: a {\it saturated} regime at high rotation rates, and an {\it unsaturated} regime for slow rotators. The transition between these two regimes can be seen to occur at $Ro \sim 0.1$.

In the discussion that follows, the rotation--activity relation is divided into these two regimes in an attempt to reveal the different physical processes at work. Approximately linear relations in log-log space are immediately apparent from this figure, despite a spread in both $R_X$ and $Ro$. This spread is likely to be caused by a number of different factors arising from the necessarily simplified analysis of the data performed here. Both the X-ray luminosity and the photometric rotation period are approximations of the real values due to the methods employed in measuring them. X-ray luminosities will vary over the course of a stellar activity cycle, by up to an order of magnitude in the case of our Sun \citep{pere00}, and will vary on shorter timescales due to the influence of coronal flares. Furthermore, the measured rotation period will be a latitudinal mean due to the unresolved nature of the stellar disk, and may also vary over the course of an activity cycle due to the varying starspot coverage.

A further source of uncertainty is due to the effects of unresolved binaries in this sample. Photometrically, the presence of a lower mass binary companion will cause the star to appear redder and more luminous, resulting in a lower mass estimate and a smaller photometric distance. Distance errors cancel out when deriving the X-ray-to-bolometric luminosity ratio, but uncertainties in the mass, and therefore spectral type, can affect a number of the parameters and quantities derived for each star. Literature spectral types were obtained for 225 of the 824 stars in the sample including 163 of the 168 binaries in our sample, reducing the influence of this effect. Another influence of binary companions is the effect on the X-ray emission, the presence of a close secondary potentially elevating the level of emission due to tidal coupling or their orbits. The 168 binaries in our samples are highlighted in Figure~\ref{allstars} and are concentrated in the unsaturated regime (which is dominated by older field stars). Their X-ray levels, in both the saturated and unsaturated regimes, all lie within the 1$\sigma$ spreads on the fits determined here, with no trend for either elevated or reduced X-ray emission. We therefore include these sources in our subsequent study of the rotation--activity relation without fear that their presence will bias the results.

\subsection{Dynamo efficiency in the unsaturated regime}
\label{s-unsat}

The unsaturated regime in the rotation--activity relation is believed to probe the efficiency of the stellar dynamo in heating the corona. \citet{pall81} found that the X-ray luminosity of solar- and late-type stars scales with projected rotational velocity, to the first order, $L_X \propto (v \mathrm{sin} i)^2$. This relationship has since been investigated by many authors, more recently in the form of the $R_X$ - $Ro$ relationship, using photometric rotation periods as well as making the distinction between stars with saturated and unsaturated X-ray emission. The low level of scatter in the $R_X - Ro$ diagram has been interpreted by many authors \citep[e.g.][]{mont01,pizz03} as evidence for a strong underlying physical relationship.

To parameterize this relationship the sample in Figure~\ref{allstars} was fitted with a two-part function of the form

\begin{equation}
R_X = \left\{ \begin{array}{ll}
  R_{X sat} & \textrm{if $Ro \leq Ro_{sat}$} \\
  C \, Ro^{\beta} & \textrm {if $Ro > Ro_{sat}$}
  \end{array} \right.
\end{equation}

\noindent
where the parameters $R_{X sat}$, $Ro_{sat}$, and $\beta$ (the X-ray luminosity ratio and Rossby number at saturation, and the power-law index) were varied to find the best fit using a $\chi^2$-minimization technique. Uncertainties on these quantities were then determined using a bootstrapping approach, iterating 1000 times and finding the standard deviation of each parameter. The parameters $R_{X sat}$ and $Ro_{sat}$ will be discussed in more detail in Section~\ref{s-sat} and were found to be log~$R_{X sat} = -3.13 \pm 0.08$ and $Ro_{sat} = 0.13 \pm 0.02$ for the best-fitting model, which has an rms scatter of $\sim$0.3~dex in $R_X$. The best-fitting slope to the unsaturated regime was found to be $\beta = -2.18 \pm 0.16$, slightly steeper than the canonical value of $\beta \simeq -2$. This fit, as shown in Figure~\ref{allstars}, over-predicts the Sun's mean X-ray luminosity by a factor of 2-3.

An alternative approach is to fit the slope of the unsaturated regime and the saturation level separately. We fitted a simple power law of the form $\mathrm{log} \, R_X = \mathrm{log} \, C + \beta \, \mathrm{log} \, Ro$ to all stars with $Ro \geq 0.2$ using the different types of linear regression fits in \citet{isob90}. We find a good agreement between the slopes derived from these different fits, suggesting that the fits are all fairly linear in the log $R_X$ -- log $Ro$ plane. We favor the Ordinary Least Squares (OLS) bisector since the objective of the fit is to estimate the underlying functional relation between the variables, as recommended by \citet{isob90}, and this method also factors in the scatter of the line in both variables. The fit gives a slope of $\beta = -2.55 \pm 0.15$ (valid in the range $0.2 < Ro < 3$, or $-3.75 > \mathrm{log} (L_X / L_{bol}) > -6.3$), significantly steeper than both the canonical value and that found from our two-part fit. \citet{mama08} fit a log-linear function to the $R_X - Ro$ distribution, with the goal of empirically deriving a correlation that would allow age estimates to be derived from X-ray luminosities (via rotation periods). Their fit does not connect with the level of saturated X-ray emission for very fast rotators but, as they note, it offers a good fit to many of the slow rotators such as the Sun.

The sample used here suffers from a number of biases due to the selection of only sources with measured X-ray fluxes and photometric rotation periods. While the biases stemming from the detectability of rotation periods are myriad and complex, the luminosity bias induced by only using sources with measured X-ray fluxes is clear. This bias will be most prominent in the unsaturated regime where X-ray luminosity ratios may reach as low as $\sim 10^{-7}$ or lower. This sample could therefore be missing some of the faintest sources at a given Rossby number, possibly resulting in a larger spread in the $R_X - Ro$ diagram than is currently observed. Such a spread could easily be induced by the increased amplitude of stellar coronal cycles that has been suggested to occur as stars age \citep[e.g.][]{mice03}. At the largest Rossby numbers it is likely that many of the faintest X-ray sources are not detected, inducing a strong bias in our sample that will affect the fits derived here.

\subsubsection{Probing the dynamo efficiency with an X-ray unbiased sample}

\begin{figure}
\begin{center}
\includegraphics[height=240pt, angle=270]{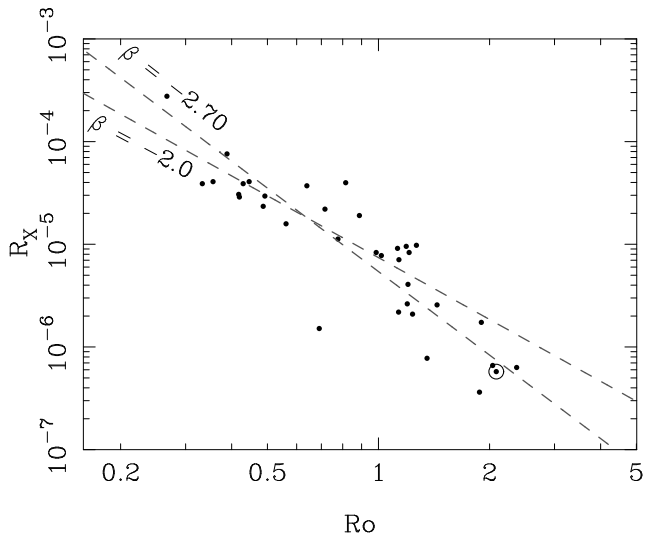}
\caption{X-ray to bolometric luminosity ratio versus Rossby number for our unbiased sample of 36 stars with unsaturated X-ray emission. The Sun is indicated with a solar symbol. The log-log OLS bisector fit, $\beta = -2.70$, is shown as a dashed line alongside a fit with the canonical slope of $\beta = -2.0$.}
\label{unbiased}
\end{center}
\end{figure}

To overcome the biases in our large sample we have attempted to compile from within our sample a smaller, X-ray unbiased sample that covers a large range in X-ray luminosity ratios and rotation periods. For this we use the list of 36 Mt. Wilson stars with rotation periods from the study by \citet{dona96}, all of which were detected by ROSAT and therefore do not suffer from X-ray luminosity biases. These 36 stars are the subsample of their entire sample of 100 observed stars with measurable rotation periods over five or more seasons. The authors discuss a number of possible biases in their sample resulting from effects such as active region growth and decay, multiple active regions, and latitudinal bands. They conclude that the resulting biases affect only $\Delta P$, not the period itself, and are either small or act to reduce the measured value of $\Delta P$. Therefore we believe that this sample of 36 stars with measured rotation periods and and X-ray luminosities is free from the majority of biases. These stars were included in our sample as part of the compilations of \citet{pizz03} and \citet{mama08}, and in Figure~\ref{unbiased} we show their distribution in the $R_X$--$Ro$ diagram, all of which fall in the unsaturated regime of coronal emission.

We fitted a simple single-part power law of the form $\mathrm{log} \, R_X = \mathrm{log} \, C + \beta \, \mathrm{log} \, Ro$ to these points, using an OLS bisector \citep{isob90}, though the slopes derived from all the different fitting methods are in good agreement. The fit gives a slope of $\beta = -2.70 \pm 0.13$ (valid in the range $0.3 < Ro < 3$, or $-4 > \mathrm{log} (L_X / L_{bol}) > -6.3$), steeper than that found from our larger sample, in agreement with our predictions of the uncertainties induced by the biases of that sample. This slope is even steeper than the canonical value of $\beta = -2$ as well as the slope found by \citet{pall81} of $\beta = -1.9 \pm 0.5$, though their use of projected rotational velocities instead of rotation periods represents a different relationship than that fitted here. However this slope is in good agreement with \citet{gude97} who derive a similar slope, $\beta = -2.64 \pm 0.12$ from a much smaller sample of 9 solar analogs comparing $L_X$ with $P_{rot}$. Our slope is derived from a larger sample of stars with a greater range of masses and incorporates the convective turnover time to compare $R_X$ with $Ro$. Such a steep slope also supports the findings of \citet{feig04} who found an unexpectedly steep decay of X-ray emission as a function of age for stars detected in the {\it Chandra} Deep Field North, which could indicate a steepening of the rotation--activity relationship. This slope also produces a much better fit to the Sun's $R_X$, over-predicting it's luminosity by only 0.2~dex.

Our fitted slope is inconsistent with the canonical value of $\beta = -2$ to a significance of $\sim$5$\sigma$. One of the implications of such a steep slope is that these observations are inconsistent with the model of a distributed dynamo operating throughout the convection zone , the efficiency of which scales as $Ro^{-2}$ \citep{noye84}. Current models of stellar dynamos (at least for solar-type stars with radiative cores) use a two-layer interface dynamo \citep[e.g.][]{park93,char97}. One of the implications of such a dynamo is that the dynamo efficiency does not scale simply as $Ro^{-2}$, but is a more complex function of a number of parameters, not all of which can be currently measured or determined from models.

Based on a scaling analysis, \citet{mont01} derived the following expression for the interface dynamo number, $N_D$

\begin{equation}
\label{e:n_d}
N_D = \frac{1}{Ro^2} \, \left[ \frac{1}{\nu} \frac{\Delta\Omega}{\Omega} \frac{r_{cz}L}{d^2} \right] \, ,
\end{equation}

\noindent
where $\nu$ is the ratio of the turbulent to magnetic diffusivities, $\Omega$ is the angular velocity at the base of the convection zone, $\Delta\Omega$ is the change in angular velocity through the tachocline, $r_{cz}$ is the radius of the base of the convective zone, $d$ is the characteristics length scale of convection at $r_{cz}$ and $L$ is the characteristic length scale of differential rotation in the tachocline. We can interpret the deviation from the proportionality $N_D \propto Ro^{-2}$ indicated by the unbiased sample of X-ray luminosity ratios illustrated in Figure~\ref{unbiased} in terms of the additional expression in $N_D$ in equations~\ref{e:n_d}. \citet{mont01} argue that the term in brackets can be further simplified by appealing to the similar scalings of $L$ and $d$. Both are proportional to the pressure scale height, $H_p$ at the convection zone base, such that

\begin{equation}
\frac{1}{\nu} \frac{\Delta\Omega}{\Omega} \frac{r_{cz}L}{d^2} \sim \frac{\Delta\Omega}{\Omega} \frac{r_{cz}}{H_p} \, .
\end{equation}

Substituting into Equation~\ref{e:n_d} we are left with a color-dependent term, $r_{cz}/H_p$, and the differential rotation in the tachocline expressed as a fraction of the angular rotation velocity, $\Delta\Omega / \Omega$. The unbiased sample used here covers only a small range in effective temperature, therefore for unsaturated magnetic activity it is reasonable to assume that $r_{cz} / H_p$ does not vary and is also independent of rotation velocity, although these factors might be responsible for some of the scatter about the mean relation in Figure~\ref{unbiased}. Recalling that the Rossby number is simply $Ro = P_{rot} / \tau_c = 2 \pi / \Omega \tau_c$, and that $\tau_c$ is approximately constant for our unbiased sample, the slope of the relation in Figure~\ref{unbiased} implies that, for solar type stars, 

\begin{equation}
N_D \propto \frac{1}{Ro^2} \frac{\Delta \Omega}{\Omega} \propto Ro^{-2.70}
\end{equation}

\noindent
and therefore that

\begin{equation}
\frac{\Delta \Omega}{\Omega} \propto \Omega^{0.7} \, ,
\end{equation}

\noindent
i.e. the fractional differential rotation in the unsaturated regime scales with the angular rotation rate approximately according to $\Omega^{2/3}$. The implication of this is that the differential rotation of solar-type stars gradually declines as they spin down.

\subsection{The saturated regime}
\label{s-sat}

Coronal X-ray saturation is known to occur for all fast-rotating solar- and late-type stars. Its onset occurs at larger rotation periods for stars of lower masses \citep{stau94}, the saturation threshold scaling with the convective turnover time, $P_{sat} \propto \tau$, such that it occurs at a constant value of the Rossby number \citep{pizz03}. The sample used here confirms this, showing little variation in $Ro_{sat}$ when the fits described above are performed in different color bins, the best fitting value for the entire sample being $Ro_{sat} = 0.13 \pm 0.02$. However, if the saturated and unsaturated regimes are fit separately (which results in a steeper slope to the unsaturated regime), the interception of these fits is shifted to $Ro_{sat} = 0.16 \pm 0.02$.

That the criteria for saturation is a parameter related to the efficiency of the stellar dynamo has been argued by some to suggest that coronal saturation is caused by saturation of the dynamo itself \citep[e.g.][]{gilm83,vilh87}, a theory supported by recent observations of magnetic flux saturation in rapidly rotating M dwarfs at Rossby numbers of $\sim$0.1 \citep{rein09}. However the saturation thresholds of other activity indicators are not so coherent. \citet{mars09} find saturation of the chromospheric Ca~{\sc ii} emission lines at $Ro \sim 0.08$, while \citet[][their Figure~15]{mama08} do not find a clear agreement between coronal and chromospheric saturation of field stars. Furthermore \citet{card07} do not find a clear saturation threshold from Mg~{\sc ii} chromospheric emission line measurements, and even argue that rotation period is an equally good diagnostic for the Mg~{\sc ii} line strength.

Other explanations for X-ray saturation have included a saturation of the filling factor of active regions on the star's surface \citep{vilh84}, motivated by a strong correlation between saturated $L_X$ and stellar radius, rather than between $L_X$ and surface temperature. However, the detection of rotational modulation in some saturated stars \citep[e.g.][]{gude95}, and the need for enhanced plasma densities to explain the observed levels \citep{wood94,drak00,test05} argues against this theory. \citet{jard99} suggested that X-ray saturation could alternatively be caused by centrifugal stripping of the corona at very fast rotation rates. However it will be shown in Section~\ref{s-evolution} that the mass-dependent functional forms of the observed saturation limit, $Ro_{sat} (M)$, and the fractional corotation radius, $r_c / R_\star (M)$, are significantly different, which argues against such a theory to the first order.

For our sample we find a mean saturation level of $R_X = -3.13 \pm 0.22$ $(1\sigma)$, almost independent of spectral type. This value is in good agreement with the two-part power law fit, though it is less well constrained. As found by previous authors \citep[e.g.][]{pizz03} we find a significantly lower mean saturation level for the highest-mass stars in our sample (F-type stars), though we will show in Section~\ref{s-evolution} that these stars are not in fact saturated, but are super-saturated and we therefore conclude that the X-ray saturation level is independent of spectral type.

\subsubsection{The saturation threshold and the angular momentum evolution of stars}

\begin{figure}
\begin{center}
\includegraphics[height=230pt, angle=270]{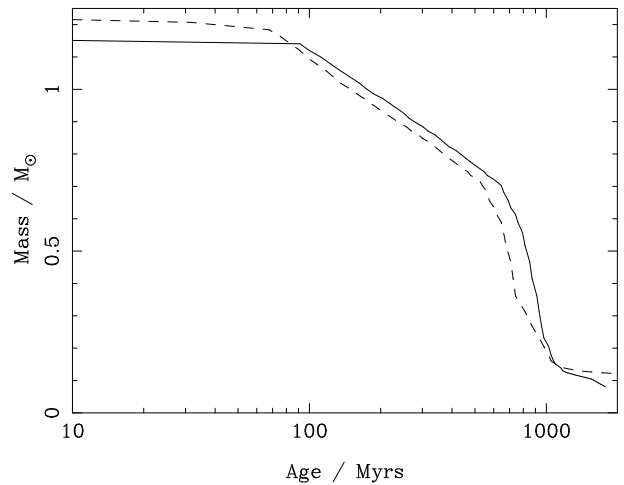}
\caption{The saturation threshold $Ro < 0.13$ in mass--age space (solid line), compared to the age at which stars transition from the rotational C sequence to the I sequence (dashed line). All ages calculated using the rotational evolution formulated by \citet{barn03} and the parameterization of \citet{mama08}.}
\label{regimes2}
\end{center}
\end{figure}

To understand the physical origin of coronal saturation, the threshold for saturation can be converted from one in terms of the Rossby number (though effectively the rotation period and the spectral type) to age using knowledge of the rotational evolution of stars. \citet{barn03} outlined an empirical formulation of the rotation period evolution of solar and late-type stars based on open cluster and field star data, characterizing two sequences in the $P_{rot} - (B-V)$ plane: the {\it convective} (C) sequence of young, fast rotators, and the {\it interface} (I) sequence of slow rotators. This empirical characterization allows ages to be derived from rotation periods for stars on the I sequence, a method known as gyrochronology. This technique results in a precision of $\pm 0.05$~dex in log~$\tau$/yr \citep[excluding absolute uncertainties in the cluster age scale,][]{mama08}. Stars on the C sequence are harder to date accurately due to star-to-star variations in the ZAMS arrival time and effects such as disk-locking in the pre-main sequence phase. Here the recent gyrochronology parameterization of \citet{mama08} has been used to convert the empirical threshold for saturation, $Ro > 0.13$, into a threshold in mass--age space, as shown in Figure~\ref{regimes2}. It is worth noting that, at the ages considered here, differences between the gyrochronological parameters of \citet{barn03}, \citet{mama08}, and \citet{meib09} are too small to be discernible in this figure.

Also shown in Figure~\ref{regimes2} is the mass-dependent age at which a star transitions from the C sequence to the I sequence. The form of these two, independently-determined, empirical transitions is remarkably similar. This implies that stars with saturated X-ray emission are found on the C sequence of fast rotators, while those with non-saturated emission can be found on the I sequence of slow rotators. A similar result to this was already hinted at from a much smaller sample by \citet{barn03b} who associated the presence of stars on the rotational I sequence with unsaturated X-ray emitters, but suggested that stars on the C sequence were coronally super-saturated and stars in the gap between the C and I sequences showed saturated X-ray emission. The association of super-saturated coronal emitters with stars on the C sequence can be argued against simply based on the very small number of super-saturated X-ray emitters (see Section~\ref{s-supersaturation}) compared to the much larger number of fast rotators found on the C sequence of young clusters.

The correlation between the transition from saturated to unsaturated coronal emission and from the rotational C sequence to the I sequence is not surprising when one considers that both angular momentum loss and coronal X-ray emission are products of the stellar magnetic dynamo and known to scale with the convective turnover time \citep[e.g.][]{mont01}. This scaling was noted by \citet{barn10a} who showed that the functional dependence of the angular momentum loss rate was related to the convective turnover time, and henceforth derived a new model for the rotational evolution of cool stars in terms of the Rossby number \citep{barn10b}. In fact the balance between coronal losses and angular momentum loss is believed to be due to the configuration of magnetic field lines at the stellar surface, with the fraction of open magnetic field lines determining the spin-down rate, while closed magnetic field lines dictate the X-ray luminosity \citep[e.g.][]{holz07}.

The similarity between two independently determined empirical transition criteria is unlikely to be a coincidence. It therefore suggests that the changes that occur within a star as it transitions from the C sequence to the I sequence are also responsible for the star leaving the saturated regime of X-ray emission. One could also consider the opposite causality, that the mechanisms responsible for the changes in X-ray emission also cause changes in the angular momentum loss rate. However, the evolution of angular momentum loss is a more complex subject that must explain other observable features (e.g. the sequence of young ultrafast rotators in clusters) and it therefore seems unlikely that the mechanism responsible for X-ray saturation is responsible for these effects as well.

The hypothesis that we are therefore left with is that the processes governing stellar angular momentum loss are also responsible for X-ray saturation. \citet{barn03} argued that the C and I sequences are due to the coupling of the stellar wind to, respectively, just the convective zone (which is decoupled from the radiative zone), and to the entire star. The transition between these zones (across the `gap' in the color--rotation period diagram) is therefore associated with the coupling of the radiative and convective zones to each other. \citet{barn03} suggests that stars on the convective sequence generate a convective or turbulent dynamo \citep[e.g.][]{durn93} and that the shear between the fast spinning radiative interior and the convective envelope eventually generates an interface dynamo that results in the transition onto the I sequence. \citet{barn10a} have since argued that the form of the rotational isochrones does not agree with the form expected for the relevant moments of inertia outlined in this model. However, the association of the two regimes with two different dynamos is worth exploring further.

Here we suggest that coronal saturation does not require a separate physical mechanism, but is a manifestation of the different dynamos that are present in stars on the convective and interface sequences. In this picture, slowly rotating stars have a cyclic, interface-type dynamo with Skumanich-style spin-down, while faster rotators have a turbulent dynamo. The efficiencies of the two different dynamos could have significantly different dependencies, in terms of stellar parameters, and therefore have different forms in the $R_X$--$Ro$ diagram. X-ray saturation would therefore not be an actual saturation in any sense, but a completely different magnetic dynamo configuration with completely different dependencies.

If this scenario is correct it partly answers the question raised by \citet{pizz03} as to why the bolometric luminosity is the only parameter necessary to determine the X-ray luminosity in the saturated regime, despite the Rossby number being considered the fundamental parameter of the efficiency of the stellar dynamo. The answer is that there are two different dynamo configurations, a convective dynamo parameterized by the bolometric luminosity, an interface dynamo parameterized by the Rossby number. This does however raise the question of why the transition between the two regimes is apparently so smooth in the $L_X / L_{bol}$ versus $Ro$ diagram. In other words, why is the luminosity ratio for unsaturated stars close to the saturation threshold equal to $\sim 10^{-3}$? The answer to this question will require a much better understanding of the two dynamo configurations present in the saturated and unsaturated regimes.

\subsection{Supersaturation}
\label{s-supersaturation}

At very high rotational velocities the fractional X-ray luminosity has been observed to decrease below the saturation level \citep{rand96}, an effect dubbed ``supersaturation''. This effect has only been persuasively observed in the coronal emission from young ($< 100$~Myrs) G and K dwarfs and is primarily based on projected rotational velocities, not rotation periods, in the young clusters IC~2391, IC~2602, and $\alpha$~Persei \citep[e.g.][]{rand96,pros96,stau97,step01} from which \citet{stau97} found a decline in X-ray saturation for stars with $Ro < 0.01$. Despite this, evidence for supersaturation in lower-mass stars is scarce; both \citet{jame00} and \citet{jeff11} could find no evidence in samples of ultra-fast rotating M dwarfs.

\begin{figure*}
\begin{center}
\includegraphics[height=480pt, angle=270]{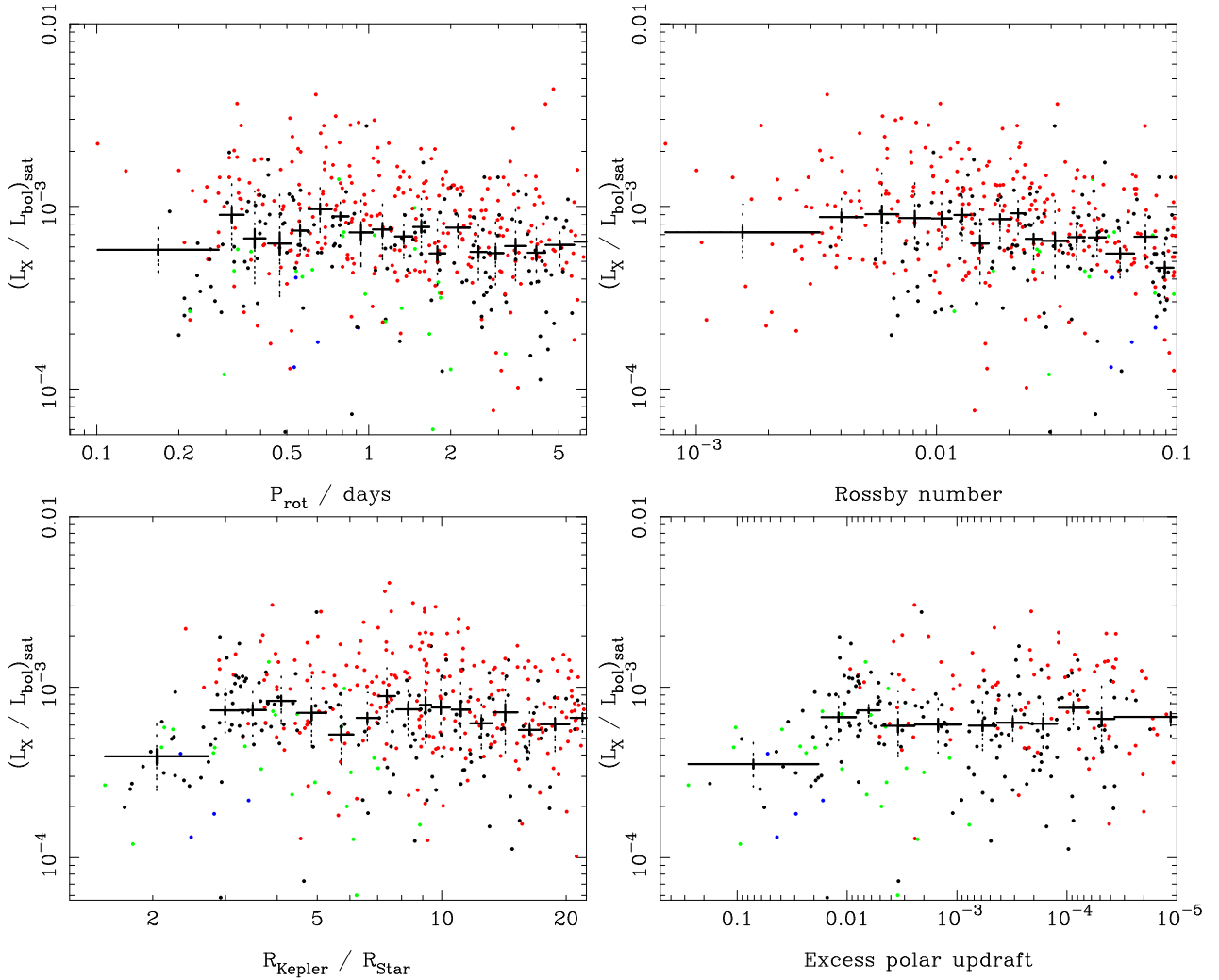}
\caption{X-ray to bolometric luminosity ratios for all stars with saturated X-ray emission ($Ro < Ro_{sat}$) as a function of the rotation period (upper left), the Rossby number (upper right), the Keplerian co-rotation radius (lower left), and the excess polar updraft acceleration (lower right). Stars are colored as per their spectral type (F-type stars are blue, G-type stars are green, K-type stars are black, and M-type stars are red). Also shown are 25-star average luminosity ratios, with standard error and standard deviation error bars in full and dotted lines respectively.}
\label{supersaturation}
\end{center}
\end{figure*}

Like coronal saturation, the phenomenon of super-saturation and its physical origin is still heavily debated. Figure~\ref{supersaturation} shows $L_X / L_{bol}$ for all stars with saturated X-ray emission as a function of various quantities suggested as parameters influencing the supersaturation effect. The first two panels of Figure~\ref{supersaturation} show the effect of rotation period and Rossby number on the X-ray luminosity ratio, neither of which show evidence for supersaturation, or a notable decline in the luminosity ratio for either the fastest rotators or the sources with the smallest Rossby numbers.

\citet{jame00} developed the theory of \citet{jard99} that the X-ray emitting volume of rapid rotators is reduced via centrifugal stripping. This theory was initially put forward to explain X-ray saturation, with the increase in coronal density balancing the decrease in volume and leading to saturation, but \citet{jame00} and \citet{jard04} used it to explain super-saturation for stars rotating fast enough that the Keplerian corotation radius becomes very close to the stellar surface. As a function of the stellar radius, the corotation radius is given by \citet{jame00}

\begin{equation}
R_{Kepler} / R_\star = \left( \frac{ G M_\star P^2} {4 \pi^2 R_\star^3} \right)^{1/3} \propto M^{1/3} \, P^{2/3} \, R_{\star}^{-1}
\end{equation}

\noindent
which, using stellar properties taken from the models of \citet{sies00}, allows the fractional X-ray luminosity to be calculated as a function of the relative corotation radius, as shown in Figure~\ref{supersaturation}. In this sample there appears to be some evidence for a decrease in the typical fractional X-ray luminosity of stars with the lowest corotation radii. Based on the mean $L_X / L_{bol}$ a decrease is seen for $R_{Kepler} / R_\star \lesssim 3$, in agreement with that found by \citet{jame00} and \citet{jeff11}. This effect is still evident if one considers only K stars, and to a lesser extent G stars, but not M stars, as found by previous authors. As parameterized in Figure~\ref{supersaturation} the bin containing the fastest rotators hints at a moderate decline to $L_X / L_{bol} = -3.41 \pm 0.26$, deviating 1$\sigma$ below the mean saturation value. A two-sample Kolmogorov-Smirnov (KS) test comparing the X-ray luminosity ratio of stars above and below the proposed corotation supersaturation threshold of $R_{Kepler} / R_\star \sim 3$ gives a probability of $< 10^{-3}$ that the two samples are drawn from the same population.

\citet{step01} offered an alternative explanation that the reduction in X-ray luminosity is caused by a decrease in the filling factor of coronal active regions on the stellar surface. They propose that a poleward migration of active regions would be induced by nonuniform heating of the convective envelope in accordance with the von Zeipel theorem. \citet{vonz24} argued that the effect of rotation on the internal radiative transfer within a star causes a higher fraction of flux at the poles than at the equator, an effect known as gravity-darkening. For sufficiently fast rotating stars this flux imbalance can also induce strong convective updrafts in the outer convective envelope \citep{step98}, capable of sweeping magnetic flux tubes in the interior to the poles before they rise to the surface.

According to the von Zeipel theorem the emergent surface flux for a rotating solid body is proportional to the local effective gravity, $g_{eff}$, given by

\begin{equation}
g_{eff} (\omega) = - \frac{GM}{R^2} + \omega^2 R sin^2 \theta
\end{equation}

\noindent
where $\omega$ is the angular rotation rate, $R$ is the radius, and $\theta$ is the colatitude \citep[differences in the theorem for a differentially rotating star are minor,][]{maed99}. \citet{step01} have argued that the resulting change in X-ray luminosity is proportional to the relative strength of this ``polar updraft'' effect at the boundary between the radiative interior and the convective envelope, which we quantify as

\begin{equation}
G_X = \frac{g_{eff} (\omega) - g (0) }{g (0)} = - \frac{ \omega^2 R_c \, sin^2 \theta }{ GM / R_c^2 } 
\end{equation}

\noindent
where $R_c$ is the radius of the radiative--convective boundary. The efficiency of this mechanism therefore scales as the rotation rate to the second power and the radius of the radiative--convective boundary to the third power. Adopting the radius of this boundary from the models of \citet[][as a function of both spectral type and age]{sies00}, Figure~\ref{supersaturation} shows the X-ray luminosity ratio as a function of the absolute value of this parameter, with a clear decline for $G_X \gtrsim 0.01$. The bin containing the stars with the strongest polar updraft shows a deviation from the mean level with a value of $L_X / L_{bol} = -3.45 \pm 0.16$, approximately 2$\sigma$ below the mean value for the saturated stars\footnote{The variance in the luminosity ratio for the supersaturated stars under the parameterization of polar updraft is smaller than that under coronal stripping, as is clear from Figure~\ref{supersaturation}. This results in a more significant deviation of the supersaturated stars from the mean value under this parameterization.}. A KS test comparing stars above and below the proposed polar updraft supersaturation level gives a probability of $< 10^{-3}$ that the two samples are drawn from the same population.

Based on the sample analyzed here, while no single theory provides a conclusive description of the supersaturation phenomenon, we can show that neither the rotation period nor the Rossby number are an accurate diagnostic of the effect, while either the Keplerian corotation radius or the excess polar updraft offer a clearer trend. Of these two, Stepien's theory of poleward migration of active coronal regions has the stronger correlation with the observed decrease in the X-ray luminosity ratio and a more significant deviation from the mean X-ray luminosity ratio saturation level.

The fundamental difference between these two theories for supersaturation concerns the underlying structure of the coronal loops that give rise to X-ray emission. Coronal stripping will occur if the loops are large, low-density, and unstable to the centrifugal force. In Figure~\ref{supersaturation} a decrease in the luminosity ratio begins at $R_{Kepler} / R_\star \sim 3$, which would suggest that coronal loops are as large as two stellar radii in height. Alternatively, coronal loops may be compact, high density, and influenced by the dynamics of surface flows that induce a polar updraft of coronal regions towards the stellar poles. The difference between the two supersaturation theories is therefore the size and density of the coronal plasma. High-resolution X-ray spectroscopy (necessary to determine such quantities) of nearby, supersaturated stars are very rare. The only such observation to date is that of the coronally supersaturated (log~$L_X / L_{bol} = -3.6$, $Ro \simeq 0.03$) star VW~Cephei by \citet{huen06} who found a coronal density of log~$N_e = 10.5 - 11.25$. For VW~Cephei this is equivalent to a coronal loop height of 0.06-0.2 stellar radii, significantly below that at which supersaturation would be expected to occur (Figure~\ref{supersaturation}) or the corotational radius of VW~Cephei itself, $R_{Kepler} / R_\star \simeq 1.75$. These observations therefore suggest that coronal loops are compact and high-density and sufficiently far below the corotation radius of the star that coronal stripping is unlikely to be responsible for supersaturation. Similar coronal densities have been observed for a sample of saturated and unsaturated stars \citep{test04}.

While it is not clear that the heating mechanism for the chromosphere is the same as those for the corona, their emission scales reasonably well and therefore chromospheric emission can offer another perspective on stellar activity. If coronal supersaturation were caused by the migration of active regions towards the poles one might expect to see supersaturation in chromospheric emission as well. \citet{mars09} and \citet{jack10} have searched for chromospheric supersaturation in a sample of F, G, K, and M-type stars but could find no evidence for a reduction in chromospheric activity indicators from a small sample of ultra-fast rotators. Since chromospherically active regions with their lower scale height should not be vulnerable to centrifugal stripping in the same way coronal active regions are, the lack of evidence for chromospheric supersaturation, if confirmed, would argue for coronal stripping as the likely mechanism for supersaturation.

The mechanism behind supersaturation remains a mystery despite this work. Coronal and chromospheric observations of even faster rotating stars that might exhibit stronger levels of supersaturation are necessary to resolve this. Such observations could provide a more significant proof of the lack of chromospheric supersaturation, or evidence for chromospheric saturation at a faster rotation rate. Increasing the sample of supersaturated stars would also allow the correlations studied here to be probed in greater detail.

\section{Coronal X-ray emission over stellar lifetimes}
\label{s-evolution}

\begin{figure}
\begin{center}
\includegraphics[height=230pt, angle=270]{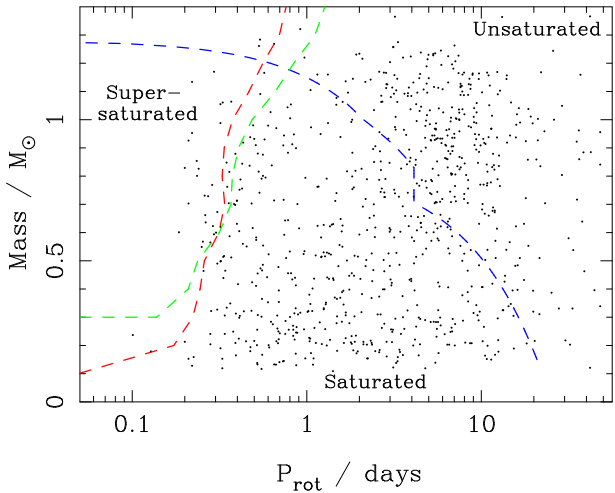}
\caption{The three regimes of coronal X-ray emission shown in Mass--$P_{rot}$ space, with all sources in this sample shown as black dots. The transition from saturated to unsaturated emission is defined by $Ro > 0.13$ (blue dashed line). The transition from saturated to supersaturated emission is shown for both the theories of coronal stripping ($r_c / R_\star \sim 3$, red dashed line) and coronal updrafts ($G_X \sim 0.01$, green dashed line).}
\label{regimes}
\end{center}
\end{figure}

In this section an attempt is made to combine the relationships and thresholds determined in this work to broadly sketch out the changes in coronal X-ray emission that occur over stellar lifetimes, as a function of the rotation period. Figure~\ref{regimes} shows the mass--rotation plane for all stars in our sample. Also shown are the borders of the three regimes of unsaturated, saturated, and super-saturated coronal X-ray emission. The super-saturated regime is defined both using the theory of coronal stripping (assuming a threshold of $r_c / R_\star \sim 3$) and the theory for polar updraft (using $G_X \sim 0.01$). The similarity between the theoretical predictions of the supersaturation thresholds of these two theories is particularly notable, despite their different theoretical formulations. This goes a long way towards explaining why both parameterizations in Figure~\ref{supersaturation} showed significant evidence for the supersaturation effect. The forms of both of the supersaturation criteria also illustrates why only G and K dwarfs have been observed to have super-saturated X-ray emission, since M dwarfs would have to be rotating exceptionally fast to exhibit the phenomenon, possibly at the break-up period, and certainly faster than the typically observed rotation periods for young, ultrafast rotators of $\sim$0.2~days.

\begin{table*}
\begin{center}
\caption{Empirical convective turnover times}
\label{results}
\begin{tabular}{cccc ccc cccc}
\tableline 
$V-K_s$ & $B-V$ & Mass ($M_\odot$) & $N_\star$ & \multicolumn{3}{c}{Fit with $\beta=-2$} &&  \multicolumn{3}{c}{Fit with $\beta = -2.70$} \\ 
\cline{5-7} \cline{9-11} 
 range & range & range & & log $R_{X sat}$ & $P_{sat}$ (days) & log $\tau$ (days) && log $R_{X sat}$ & $P_{sat}$ (days) & log $\tau$ (days) \\ 
\tableline
1.14 -- 1.48  &  0.46 -- 0.61 & 1.16 -- 1.36 &  81  & $-4.16^{+0.42}_{-0.14}$  &  $0.56^{+0.10}_{-0.28}$  &  $1.02^{+0.04}_{-0.07}$  &&  $-4.26^{+0.12}_{-0.10}$  &  $0.70^{+0.08}_{-0.08}$  &  $1.01^{+0.05}_{-0.04}$ \\ 
1.48 -- 1.79  &  0.61 -- 0.75 & 1.02 -- 1.16 &  84  & $-3.28^{+0.08}_{-0.08}$  &  $0.24^{+0.04}_{-0.04}$  &  $1.14^{+0.02}_{-0.03}$  &&  $-3.34^{+0.08}_{-0.10}$  &  $0.42^{+0.04}_{-0.04}$  &  $1.08^{+0.02}_{-0.02}$ \\ 
1.81 -- 2.22  &  0.76 -- 0.92 & 0.89 -- 1.02 &  79  & $-3.36^{+0.12}_{-0.06}$  &  $0.40^{+0.04}_{-0.06}$  &  $1.26^{+0.02}_{-0.03}$  &&  $-3.44^{+0.12}_{-0.06}$  &  $0.56^{+0.02}_{-0.06}$  &  $1.18^{+0.02}_{-0.02}$ \\ 
2.23 -- 2.80  &  0.92 -- 1.12 & 0.77 -- 0.89 &  79  & $-3.26^{+0.04}_{-0.04}$  &  $0.52^{+0.02}_{-0.04}$  &  $1.43^{+0.02}_{-0.03}$  &&  $-3.24^{+0.02}_{-0.06}$  &  $0.62^{+0.04}_{-0.02}$  &  $1.32^{+0.02}_{-0.02}$ \\ 
2.81 -- 3.34  &  1.13 -- 1.31 & 0.63 -- 0.77 &  84  & $-3.16^{+0.04}_{-0.04}$  &  $0.58^{+0.08}_{-0.04}$  &  $1.54^{+0.06}_{-0.04}$  &&  $-3.20^{+0.06}_{-0.04}$  &  $0.70^{+0.04}_{-0.06}$  &  $1.41^{+0.04}_{-0.05}$ \\ 
3.36 -- 3.68  &  1.32 -- 1.41 & 0.47 -- 0.62 &  81  & $-3.10^{+0.06}_{-0.06}$  &  $0.68^{+0.04}_{-0.06}$  &  $1.67^{+0.03}_{-0.05}$  &&  $-3.10^{+0.04}_{-0.08}$  &  $0.74^{+0.08}_{-0.04}$  &  $1.49^{+0.07}_{-0.03}$ \\ 
3.69 -- 4.19  &  1.41 -- 1.49 & 0.26 -- 0.47 &  80  & $-3.12^{+0.06}_{-0.04}$  &  $0.84^{+0.08}_{-0.08}$  &  $1.82^{+0.06}_{-0.06}$  &&  $-3.14^{+0.04}_{-0.04}$  &  $0.98^{+0.10}_{-0.08}$  &  $1.71^{+0.10}_{-0.06}$ \\ 
4.21 -- 4.62  &  1.50 -- 1.55 & 0.18 -- 0.25 &  83  & $-3.02^{+0.04}_{-0.06}$  &  $0.90^{+0.28}_{-0.08}$  &  $1.93^{+0.27}_{-0.08}$  &&  $-3.08^{+0.08}_{-0.02}$  &  $1.18^{+0.22}_{-0.28}$  &  $1.94^{+0.21}_{-0.27}$ \\ 
4.63 -- 4.93  &  1.55 -- 1.60 & 0.14 -- 0.18 &  84  & $-3.10^{+0.04}_{-0.02}$  &  $1.22^{+0.04}_{-0.24}$  &  $2.21^{+0.04}_{-0.22}$  &&  $-3.10^{+0.04}_{-0.02}$  &  $1.22^{+0.04}_{-0.14}$  &  $1.97^{+0.04}_{-0.14}$ \\ 
4.95 -- 6.61  &  1.61 -- 1.95 & 0.09 -- 0.14 &  79  & $-3.12^{+0.04}_{-0.04}$  &  $1.34^{+0.04}_{-0.04}$  &  $2.32^{+0.03}_{-0.05}$  &&  $-3.12^{+0.02}_{-0.04}$  &  $1.38^{+0.04}_{-0.02}$  &  $2.12^{+0.03}_{-0.02}$ \\ 
\tableline 
\end{tabular} 
\newline
Empirical convective turnover times fitted from the data divided into $V-K_s$ color bins (with equivalent $B-V$ color and main sequence mass bins also shown). Fits were performed assuming a constant saturation limit, $Ro = 0.13$, as determined above, and the slope of the unsaturated rotation -- activity regime to be either $\beta = -2$ (the standard value) or $\beta = -2.70$, as estimated from the fit to to our unbiased sample. 1$\sigma$ uncertainties are also shown, estimated using a bootstrapping technique.
\end{center} 
\end{table*} 

Also clear from Figure~\ref{regimes} is that the majority of F-type stars ($M \gtrsim 1.15 M_\odot$) will never exhibit saturated X-ray emission over their lifetimes, passing from the super-saturated regime straight to the unsaturated regime. This explains why previous authors \citep[e.g.][]{pizz03}, as well as this study (see Section~\ref{s-tau}), find a lower X-ray luminosity ratio for saturated F-type stars, $(L_X / L_{bol})_{sat} \simeq -4.3$, than for solar and lower-mass stars, $(L_X / L_{bol})_{sat} \approx -3.2$. Since they are not truly in the saturated regime their emission should not be treated as such.

The theoretical form of constant corotation radius in the mass--rotation diagram is also notably different from the empirically-derived form of the saturation boundary, $Ro \sim 0.13$, which suggests that centrifugal stripping is unlikely to be responsible for X-ray saturation\footnote{A caveat to this is that X-ray saturation via centrifugal stripping is suggested to be a balance between the decreasing coronal volume (as the corotation radius decreases) and an increasing plasma density (as rotation rate increases) that results in an apparently uniform saturation level \citep{jard99}, therefore the comparison of these two curves does not represent the complete picture.}.

\section{Empirical determination of the convective turnover time}
\label{s-tau}

It is clear from the work performed here that a fundamental quantity in the rotation--activity relation and the evolution of angular momentum loss for all cool stars is the convective turnover time, $\tau$ (e.g. Figure~\ref{allstars}). This color-dependent quantity can be estimated theoretically \citep[e.g.][]{kim96,vent98} or determined empirically \citep[e.g.][]{noye84,step94}, though both methods have mainly been applied to solar-type stars, with very little work on very low mass stars. \citet{pizz03} were the first to extend the empirical approach into the M-dwarf regime, but even their study was limited to only 21 stars with $B-V > 1.55$. The sample used here is almost an order of magnitude larger than that used by \citet{pizz03}, which contained 164 stars with $B-V > 1.55$ and 26 stars with $B-V > 1.7$, and could therefore offer a considerably improved empirical determination of $\tau$.

Following \citet{step94} we derive an empirical estimate of $\tau$ by assuming that the $L_X / L_{bol} - R_e$ diagram can be divided into saturated and non-saturated regimes with the formulations described above. To do this we have divided our sample into 10 bins in $V-K_s$ or $B-V$ color, with an approximately equal number of stars in each bin (see Table~\ref{results}) and then fitted the equations

\begin{equation}
\frac{L_X}{L_{bol}} = \left\{ \begin{array}{ll} 
	C_{B-V} \, P_{rot}^{\beta} & \textrm{for} P_{rot} > P_{sat} \\
	\left( \frac{L_X}{L_{bol}} \right)_{sat} &  \textrm{for} P_{rot} \leq P_{sat}
\end{array} \right.
\end{equation}

\noindent
varying the parameters $(L_X / L_{bol})_{sat}$ and $P_{sat}$ to provide the best fit. This was performed using $\beta$ values of -2.0 (the canonical value) and -2.70 (the value determined here). From these fits the color-dependent constant, $C_{V-K_s} = C \tau^2$, may be determined and, therefore, the color-dependence of the convective turnover time, $\tau$, by setting the scaling constant, $C$, so that our values of $\tau$ for solar-mass stars match those of \citet{noye84} for the Sun. The results of this fitting process and the empirical determinations of the convective turnover time are listed in Table~\ref{results}, with 1$\sigma$ confidence intervals determined using a bootstrapping technique. Also shown are the main-sequence masses of each bin, converting the $V-K_s$ color to mass using the models of \citet{sies00}.

\begin{figure}
\includegraphics[height=240pt, angle=270]{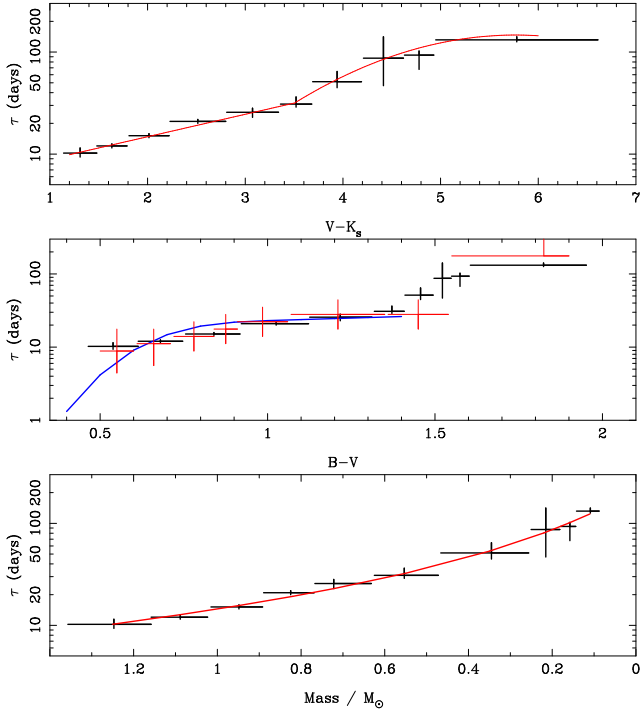}
\caption{The empirically-determined convective turnover time as a function of $V-K_s$ color (top), $B-V$ color (middle), and main-sequence mass (bottom), using the best-fitting model determined above ($\beta = -2.70$) and with 1$\sigma$ error bars for all points. In the top panel the two-part power-law fit described in the text is shown with a red line. In the middle panel our values are compared with the empirical determinations of \citet[][blue line]{noye84} and \citet[][red points]{pizz03}. In the bottom panel a second order polynomial fit as a function of log~(M) is also shown. Note that the x-axis of the bottom panel has been reversed to mimic the color-scale of the other two panels.}
\label{tau}
\end{figure}

Figure~\ref{tau} shows a comparison of our values with previous empirical estimates from the literature. Our new empirical values are in good agreement with previous estimates but have smaller error bars. At the high-mass end we do not observe the turnover towards lower convective turnover times found by \citet{noye84}, though their data only extended to $B-V \sim 0.45$ and are extrapolated beyond that. As shown in Section~\ref{s-evolution}, the fact that late F-type stars may never pass through the saturated regime will affect the determination of convective turnover times with this method and so also the validity of both our results and previous methods that ignored the effects of supersaturation.

At low masses our results show a rise in the convective turnover time as found by \citet{pizz03}. This rise occurs around $B-V \sim 1.4$, or spectral type M0, significantly before the fully-convective boundary that occurs at types M3-M4 ($B-V \sim 1.5-1.7$). Unlike previous studies we can resolve this rise and find that it begins to plateau again around $B-V \sim 1.6$, which is closer to the point at which stars are believed to become entirely convective and where it has been suggested a different dynamo mechanism may be at work. However, if a different dynamo mechanism is at work, then the parameterization used here to calculate the convective turnover time may not be accurate and therefore these values, and the entire use of the convective turnover time for such stars, may be invalid.

The rise in $\tau$ seen at $B-V \sim 1.4$ is less evident when one considers the convective turnover time as a function of the $V-K_s$ color (Figure~\ref{tau}), where the relationship between the two quantities is almost exponential. An OLS (Y$|$X) fit to log~$\tau$ as a function of $V-K_s$ provides a simple relation $\textrm{log} \, \tau = 0.64 + 0.27 \, (V-K_s)$ with an rms dispersion in $\tau$ of 0.058~dex. However a discontinuity is still visible at $V-K_s \sim 3.75$ (type M0) and therefore a two-part power-law is more appropriate. This was fit as

\begin{equation}
\textrm{log} \, \tau = \left\{ \begin{array}{ll}
  0.73 + 0.22 \, X & \textrm{for} X < 3.5\\
  -2.16 + 1.50 \, X - 0.13 \, X^2   & \textrm {for} X > 3.5
  \end{array} \right.
\end{equation}

\noindent
where $X = V-K_s$; the fit is valid over the range $1.1 < V-K_s < 6.6$. This two-part fit has an rms dispersion in log~$\tau$ of 0.031~dex, an improvement of a factor of two over the single power-law fit.

The different appearance between the two parameterizations as a function of color in Figure~\ref{tau} is almost certainly due to the different dependencies of the two colors on whatever is the relevant physical parameter that determines the convective turnover time. In Figure~\ref{tau} we also show the convective turnover time as a function of main-sequence stellar mass using the 1~Gyr stellar isochrones from \citet{sies00}. Under this parameterization the convective turnover time shows a much smoother rise towards lower masses with no evidence of the discontinuity seen as a function of the two colors. This could suggest that the stellar mass is the relevant physical parameter that determines the convective turnover time, or at least is more closely related to it than the two colors that have complex and non-linear dependencies on most stellar parameters.

An OLS (Y$|$X) fit gives the linear relation $\textrm{log} \, \tau = 2.12 - 0.96 \, (M / M_\odot)$, with an rms of 0.060~dex, but a much better fit can be obtained using a second order log--log polynomial

\begin{equation}
\textrm{log} \, \tau = 1.16 - 1.49 \textrm{log} (M / M_\odot) - 0.54 \textrm{log}^2 (M / M_\odot)
\end{equation}

which has an rms dispersion in log~$\tau$ of 0.028, the lowest of any of the fits here, and valid over the range $0.09 < M / M_\odot < 1.36$. 

It should be noted that the validity of any of these relationships are still highly uncertain for $V-K_s > 5$ ($B-V > 1.6$, $M < 0.15 M_\odot$) due to the paucity of data for low-mass stars. Despite this, these are certainly the most accurate and comprehensive empirical estimates of the convective turnover time for solar- and late-type stars to date. Observations of slowly rotating, X-ray unsaturated M dwarfs will be necessary to improve these but are, unfortunately, very difficult to obtain \citep[but see e.g.][]{robr09}.

\section{Conclusions}

The largest sample to date of solar and late-type stars with well measured X-ray luminosities and photometric rotation periods has been gathered from the literature. This sample of 824 stars is over three times larger than previous samples and is used here to study the relation between rotation (in the form of the Rossby number, $Ro = P_{rot} / \tau$) and stellar activity (in the form of the X-ray luminosity ratio, $R_X = L_X / L_{bol}$). The $R_X - Ro$ plane is divided into the unsaturated, saturated, and super-saturated regimes of coronal X-ray emission, and a number of fits are performed on these regimes.

In the unsaturated regime, where the relationship between $Ro$ and $R_X$ probes the efficiency of the stellar dynamo, we use a smaller unbiased sample of 36 stars to fit a slope of $\beta = -2.70 \pm 0.13$, significantly steeper than the previously assumed value of $\beta = -2$. Such a steep slope is inconsistent with a distributed dynamo model and lends weight to the theory of an interface dynamo \citep{park93} and an additional term in the dynamo number equation. Using the scaling analysis of \citet{mont01} for the interface dynamo number, we show that this result implies $\Delta \Omega / \Omega \propto \Omega^{0.63}$ for solar-type stars, i.e., that the fractional differential rotation in the unsaturated regime scales with the angular rotation rate to the power of two thirds. The steeper relationship between rotation rate and activity level also provides a better fit to the Sun's activity level than the canonical $\beta = -2$ slope that over-fitted the Sun's X-ray luminosity by an order of magnitude.

In the saturated regime we find a mean saturation level of $R_X = -3.13 \pm 0.22$, which we argue is independent of spectral type once the super-saturated F-type stars are removed. We explore evidence for the super-saturation phenomenon by parameterizing the activity level as a function of a number of parameters suspected of causing supersaturation. We find that the parameters of rotation period and Rossby number do not show a significant correlation with activity level, while the parameters of the corotation radius and the excess polar updraft both show a significant deviation from the mean saturation level. Two sample KS tests comparing saturated and super-saturated stars (using both parameters to separate the two populations) find a negligible probability that they are drawn from the same population. The theory of polar updraft suggested by \citet{step01} is found to provide the most significant deviation from the mean saturation level, and combined with other observations we argue that this is the more likely explanation for the supersaturation phenomenon. 

The color-dependent form of both supersaturation theories are shown to be significantly different from the empirically-determined saturation criteria, such that it is unlikely that either are responsible for the saturation of coronal emission, which we find to occur at a Rossby number of $Ro = 0.13$. Instead, based on the coincidence between the empirically determined saturation threshold and the time at which stars transition from the rotational convective sequence to the interface sequence \citep{barn03}, we suggest that the phenomenon of dynamo saturation is actually caused by a change in the dynamo configuration of the star. This theory precludes the need for an actual saturation effect and argues that the two main regimes of coronal emission should be treated as separate probes of different dynamo configurations. 

Finally this sample is used to determine a new estimate of the empirical convective turnover time as a function of the $B-V$ color, the more preferable $V-K_s$ color, and the main-sequence stellar mass. When parameterized as a function of either photometric $B-V$ color, the convective turnover time shows an increase around spectral-type M0, notably before the fully convective boundary ($\sim$M3-M4). A two-part power-law fit to the convective turnover time as a function of $V-K_s$ color provides the simplest method for estimating this parameter with readily available photometry. When parameterized as a function of stellar mass, the convective turnover time shows a much smoother rise towards lower masses with no evidence for a discontinuity, possibly hinting at the relevant physical parameter. The convective turnover time can be well fit as a second order log--log polynomial.

\acknowledgments

We would like to thank the anonymous referee for a prompt and helpful report that improved this work. This work has made use of NASA's Astrophysics Data System, the Simbad and Vizier databases (operated at CDS, Strasbourg, France) and data from the {\it ROSAT}, {\it XMM-Newton}, and {\it Chandra} X-ray Observatories. This work was funded by {\it Chandra} grant AR9-0003X. JJD was supported by NASA contract NAS8-39073 to the {\it Chandra} X-ray Center (CXC) during the course of this research and thanks the CXC director, Harvey Tananbaum, and the CXC science team for advice and support. GWH acknowledges support from NASA, NSF, Tennessee State University, and the State of Tennessee through its Centers of Excellence program.

\begin{appendix}

\section{APPENDIX A : Observations of FEPS stars}

We acquired time-series photometric observations of 100 stars drawn from the {\it Spitzer} Legacy Science Program Formation and Evolution of Planetary Systems \citep[FEPS,][]{meye06}. Nightly photometric observations of these stars were acquired with several of the Automatic Photometric Telescopes (APTs) operated by Tennessee State University (TSU) and located at Fairborn Observatory in southern Arizona. The telescopes included the T2 0.25~m, the T3 0.40~m, the T4 0.75~m, and the T10, T11, and T12 0.80~m APTs. Each FEPS star was observed differentially with respect to two or three comparison stars; all observations were made through standard filters and transformed to the Johnson $BVRI$ or to the Str\"omgren $by$ photometric system. Further information about the design and operation of the APTs as well as the observing and data reduction techniques and the resulting photometric precision can be found in \citet{henr95b,henr95,henr99b} and \citet{eato03}. 

Most of the FEPS stars were observed once each clear night throughout one of their individual observing seasons between 2002 September and 2004 July. A few of the stars have been observed for a decade or more because they are part of an ongoing, long-term program to monitor subtle brightness changes in solar-type stars \citep[see, e.g.,][]{henr99b,hall09}. All data sets were subjected to periodogram analysis based on least-squares fitting of sine curves over a range of trial periods.  Definitive photometric periods were determined for 72 of the 100 FEPS targets, of which 65 also have well-constrained X-ray luminosities and appear in our catalog. We take each photometric period to be the stellar rotation period made visible by rotational modulation of the visibility of photospheric starspots. \citet{henr95c} provide several dozen examples of low-amplitude light curves of active cool stars discovered with these same techniques.

\section{APPENDIX B : The Activity-Rotation Catalog, naming nomenclature and references}

The final catalog of 824 sources with X-ray luminosities and rotation periods is available online. The catalog includes a number of stellar parameters derived from the available photometry and the models of \citet{sies00}. For the cluster stars these are dependent on the properties of the cluster itself (see Table~\ref{sources}). To allow the catalog to be easily cross-referenced with other works, source names are provided for sources, where available, from the Henry Draper catalog \citep[HD,][186 sources]{cann18}, the Hipparcos catalog \citep[HIP,][22 sources]{perr97}, the Catalog of Nearby Stars \citep[GJ,][37 sources]{glie91}, the General Catalog of Variable Stars \citep[GCVS, Version 2011 Jan,][293 sources]{samu09}, and the 2MASS catalog (435 sources). The default name is taken from the X-ray or rotation period paper in the case of cluster stars, and for fields stars is chosen according to the catalog order listed above. 202 sources remain without a name.

The final catalog will be made available online. For convenience we include here only a list of the different fields in the catalog (Table~\ref{columns}) and an explanation of their meanings. References for all the X-ray luminosities, rotation periods, and spectral types used in this work are also listed as part of Table~\ref{columns}.

\begin{table}
\begin{center}
\caption{List of fields in the Activity-Rotation catalog} 
\label{columns}
\begin{tabular}{lll}
\tableline 
Label			& Units		& Explanation \\
\tableline
Name			&			& Default source name \\
HD number		&			& Henry Draper catalog number\\
HIP number		&			& Hipparcos catalog number \\
GJ number		&			& Catalog of Nearby Stars number \\
GCVS name		&			& General Catalog of Variable Stars name \\
2MASS name		&			& 2MASS source name \\
RA				& deg		& Right ascension (J2000) \\
Dec				& deg		& Declination (J2000) \\
Cluster			& 			& Name of cluster source belongs to (see Table~\ref{sources}) \\
Distance			& pc			& Distance \\
$V$				& mag		& Cousins V-band magnitude \\
$V-K_s$			& mag		& Dereddened ($V-K_s$) color\\
SpT				& 			& Spectral type (if used to infer a dereddened $V-K_s$ color) \\
r\_SpT			&			& Reference for spectral type (see list below) \\
log $L_X$			& erg s$^{-1}$	& Logarithm of the X-ray luminosity \\
r\_$L_X$			& 			& Reference for the X-ray data (see list below) \\
$P_{rot}$			& days		& Rotation period \\
r\_$P_{rot}$		& 			& Reference for the rotation period (see list below) \\
$M_\star$			& M$_\odot$	& Stellar mass from Siess models \\
$R_\star$			& R$_\odot$	& Stellar radius from Siess models \\
$L_\star$			& L$_\odot$	& Logarithm of stellar luminosity from Siess models \\
$T_{eff}$			& K			& Stellar effective temperature \\
$d_{cz} / R$		&			& Height of radiative-convective boundary from Siess models (as a function of the stellar radius) \\
log $L_X / L_{bol}$	&			& Logarithm of X-ray to bolometric luminosity \\
\tableline
\end{tabular} 
\end{center}
{\bf References for spectral types:} B46 = \citet{binn46}, B84 = \citet{breg84}, B85 = \citet{bide85}, B95 = \citet{bali95}, B97 = \citet{bouv97}, C18 = \citet{cann18}, D11 = \citet{delo11}, G01 = \citet{gray01}, G03 = \citet{gray03}, G06 = \citet{gray06}, H56 = \citet{heck56}, H62 = \citet{herb62}, H99 = \citet{houk99}, H02 = \citet{henr02}, J74 = \citet{joy74}, K67 = \citet{kraf67}, K89 = \citet{keen89}, L82 = \citet{lu82}, M65 = \citet{morg65}, M81 = \citet{mund81}, R41 = \citet{ramb41}, R66 = \citet{rebe66}, S51 = \citet{schw51}, S86 = \citet{step86}, T06 = \citet{torr06}, U72 = \citet{upgr72}, W48 = \citet{wils48}, W63 = \citet{wils63}. \newline
{\bf Rerences for X-ray luminosities:} B03 = \citet{brig03}, F03 = \citet{fran03}, J03 = \citet{judg03}, J11 = \citet{jeff11}, M90 = \citet{mice90}, M96 = \citet{mice96}, M99 = \citet{mice99}, P03 = \citet{pizz03}, P06 = \citet{pill06}, R95 = \citet{rand95}, S94 = \citet{stau94}, S04 = \citet{schm04}, V99 = \citet[][ROSAT all-sky bright source catalog]{voge99}. \newline
{\bf References for rotation periods:} B96 = \citet{bali96}, B97 = \citet{bouv97}, C99 = \citet{cuti99}, D96 = \citet{dona96}, D11 = \citet{delo11}, FEPS = FEPS Program (see Appendix), GE06 = \citet{ge06}, H95 = \citet{henr95c}, H00 = \citet{henr00}, H06 = \citet{hebr06}, H10 = \citet{hart10}, H11 = \citet{hart11}, I07 = \citet{irwi07}, I08 = \citet{irwi08}, K02 = \citet{koen02}, K07 = \citet{kira07}, M10 = \citet{mess10}, N07 = \citet{nort07}, P03 = \citet{pizz03}, P04 = \citet{paul04}, P05 = \citet{pojm05}, Q01 = \citet{quel01}, S00 = \citet{stra00}, S11 = \citet{scho11}, W98 = \citet{wich98}, X06 = \citet{xing06}, X07 = \citet{xing07}. \newline
\end{table}

\end{appendix}

\bibliography{/Users/nwright/Documents/Work/tex_papers/bibliography.bib}

\end{document}